\documentclass[prd, amsfonts, twocolumn, nofootinbib, showpacs]{revtex4}
\usepackage{graphicx, epsfig}
\usepackage{color}
\usepackage{amsmath}
\newcommand{\be}{\begin{equation}}
\newcommand{\ee}{\end{equation}}
\newcommand{\bea}{\begin{eqnarray}}
\newcommand{\eea}{\end{eqnarray}}

\newcommand{\gapp}{\mathrel{\raise.3ex\hbox{$>$}\mkern-14mu \lower0.6ex\hbox{$\sim$}}}
\newcommand{\lapp}{\mathrel{\raise.3ex\hbox{$<$}\mkern-14mu \lower0.6ex\hbox{$\sim$}}}
\def\bbox{{\,\lower0.9pt\vbox{\hrule \hbox{\vrule height 0.2 cm
\hskip 0.2 cm \vrule  height 0.2 cm}\hrule}\,}}

\begin{document}
\title{A note on the Hamiltonian of the real scalar field}
\author{De-Chang Dai}
\affiliation{ Astrophysics, Cosmology and Gravity Centre, University of Cape Town, Rondebosch, Private Bag, 7700, South Africa}

\begin{abstract}
\widetext

We address the question of ambiguity in defining a Hamiltonian for a scalar field. 
We point out that the Hamiltonian for a real Klein-Gordon scalar field must be consistent with the energy density obtained from the Schrodinger equation in the non-relativistic regime. To achieve this we had to add some surface terms (total divergencies) to the standard Hamiltonian, which in general will not change the equations of motion, but will redefine energy. As an additional requirement, a Hamiltonian must be able to reproduce the equations of motion directly from Hamilton's equations defined by the principle of the least action. We find that the standard Hamiltonian does not always do so and that the proposed Hamiltonian provides a good non-relativistic limit. This is a hint that something as simple as the Hamiltonian of the real Klein-Gordon scalar field has to be treated carefully. We had illustrated our discussion with an explicit example of the kink solution,

\end{abstract}


\pacs{}
\maketitle

\section{introduction}
The Klein-Gordon equation was first proposed as a relativistic version of the Schrodinger equation, but it contained two fundamental problems at that time. The first problem was that the negative energy solutions were predicted. The negative energy solutions were at first considered unphysical, however, it turned out that they actually predicted the existence of  antiparticles \cite{Dirac:1928hu}.

The other problem was that the wavefunction, $\psi$, does not allow for $|\psi|^2$ to be probability density, since $\psi^* \partial_t \psi -\psi \partial_t\psi^*$ may lead to negative probability. This problem turns out to be creation and annihilation of particles in the quantum physics \cite{Pauli:1934xm}. At the same time the energy of the field is not directly from the angular frequency, $k^0$, but from the Hamiltonian $H=\int d^3x (\partial_t\psi)^2/2+(\partial_x\psi)^2/2+m^2\psi^2/2$.

This Hamiltonian comes from the Lagrangian, which is predicted from the Klein-Gordon equation. However, it is well known that the Lagrangian is not unique in classical mechanics. Each Lagrangian generates its own Hamiltonian and therefore, the Hamiltonian is also not unique. In order to have a self-consistent result, the Hamilton equations generated by the Hamiltonian must be an equation of motion. This provides a new constraint which limit the degree of freedom of the Hamiltonian and possibly the Lagrangian. Furthermore, the Hamiltonian is considered to be the energy of a system and Hamiltonian density is considered to be energy density. Therefore, by considering a non-relativistic approximation, Schrodinger's quantum physics, one can decide which Hamiltonian is the correct one. 

In this following sections, we first review the real Klein-Gordon equation, and then use the same strategy to find out possible Hamiltonians. We then compare the Hamiltonian with the energy density predicted by the Schrodinger equation. It turns out the suitable Hamiltonian in this case is not $H=\int d^3x (\partial_t\psi)^2/2+(\partial_x\psi)^2/2+m^2\psi^2/2$, but rather $H=\int d^3x (\partial_t\psi)^2/2-\psi\partial_x^2\psi/2+m^2\psi^2/2$. This discrepancy is a hint that the Hamiltonian of a real Klein-Gordon field is needed. We then check if the  Hamiltonians will generate the correct equation of motion. We will show that the new Hamiltonian generates the correct equation of motion, but the standard one does not. Generating the correct equation of motion provides an indicator which is able to confirm which Hamiltonian is the right one. A similar problem has been pointed out in the linear tension wave \cite{Burko}. They found a similar replacement is needed for the linear-tension-wave Hamiltonian.

As a consequence, all the models of scalar fields where energy is important must be reconsidered. For example scalar-tensor gravity models widely used in cosmology will have different energy density from those typically considered.  

Throughout the paper we focus on a real scalar field only. The discussion presented here may apply to a complex field. However, this is not the subject of this paper.

\section{Review on a Klein-Gordon equation}
Let's consider the 2D Klein-Gordon equation
\begin{equation}
\label{motion}
\partial^2_t \psi-\partial_x^2 \psi+m^2\psi =0
\end{equation}
This equation describes the motion of a mass $m$ free particle. We multiply this equation with $\partial_t \psi$ and re-arrange it to get

\begin{equation}
\label{continue1}
\partial_t \Big(\frac{1}{2}(\partial_t \psi)^2+\frac{1}{2}(\partial_x \psi)^2+\frac{1}{2}m^2 \psi^2\Big)-\partial_x (\partial_t\psi \partial_x \psi)=0
\end{equation}

We now multiply it by $\partial_x \psi$ and get

\begin{equation}
\label{continue2}
-\partial_t (\partial_t\psi \partial_x \psi)+\partial_x \Big(\frac{1}{2}(\partial_t \psi)^2+\frac{1}{2}(\partial_x \psi)^2-\frac{1}{2}m^2 \psi^2)\Big)=0
\end{equation}

The usual continuity equation is

\begin{eqnarray}
\partial_t T^{t t}+\partial_x T^{t x}&=&0\\
\partial_t T^{x t}+\partial_x T^{x x}&=&0
\end{eqnarray}

By comparing, we expect

\begin{eqnarray}
\label{tensor}
T^{tt}&=&\frac{1}{2}(\partial_t \psi)^2+\frac{1}{2}(\partial_x \psi)^2+\frac{1}{2}m^2 \psi^2\\
T^{tx}&=& T^{xt}=-\partial_t \psi \partial_x \psi\\
T^{xx}&=&\frac{1}{2}(\partial_t \psi)^2+\frac{1}{2}(\partial_x \psi)^2-\frac{1}{2}m^2 \psi^2
\end{eqnarray}

This implies that the Hamiltonian density is

\begin{equation}
\mathcal{H}=T^{tt}=\frac{1}{2}(\partial_t \psi)^2+\frac{1}{2}(\partial_x \psi)^2+\frac{1}{2}m^2 \psi^2
\end{equation}

and the Lagrangian density is

\begin{equation}
\mathcal{L}=\frac{1}{2}(\partial_t \psi)^2-\frac{1}{2}(\partial_x \psi)^2-\frac{1}{2}m^2 \psi^2
\end{equation}

This Lagrangian gives the Klein Gordon equation by following the standard procedure.

\section{Other Hamiltonians for the same equations of motion}
Although equation (\ref{continue1}) is the standard one and is believed to be the continuity equation of the field, there are other equations that are of the same form. If one re-arranges equation (\ref{continue1}) and (\ref{continue2}) differently, one has (for some other solutions please see the appendix)

\begin{equation}
\partial_t\Big(\frac{1}{2}(\partial_t\psi )^2-\frac{1}{2}\psi \partial_x^2 \psi+\frac{1}{2}m^2\psi^2 \Big)-\partial_x\Big(\frac{1}{2}\partial_x\psi \partial_t \psi-\frac{1}{2}\psi\partial_x\partial_t \psi\Big)=0
\end{equation}

and

\begin{eqnarray}
&-&\partial_t\Big(\frac{1}{2}\partial_t\psi \partial_x \psi-\frac{1}{2}\psi\partial_x\partial_t \psi \Big)\nonumber \\
&+&\partial_x\Big(-\frac{1}{2}\psi \partial_t^2\psi+\frac{1}{2}(\partial_x\psi)^2 -\frac{m^2}{2}\psi^2\Big)=0
\end{eqnarray}

If these two are interpreted as the continuity equations, then
\begin{eqnarray}
T^{tt}&=&\frac{1}{2}(\partial_t\psi )^2-\frac{1}{2}\psi \partial_x^2 \psi+\frac{1}{2}m^2 \psi^2\\
T^{tx}&=& T^{xt}=-\frac{1}{2}\partial_t\psi \partial_x \psi+\frac{1}{2}\psi\partial_x\partial_t \psi\\
T^{xx}&=&-\frac{1}{2}\psi \partial_t^2\psi+\frac{1}{2}(\partial_x\psi)^2 -\frac{m^2}{2}\psi^2
\end{eqnarray}

 Therefore, the new Hamiltonian density may be in the form

\begin{equation}
\mathcal{H}'=T^{tt}=\frac{1}{2}(\partial_t\psi )^2-\frac{1}{2}\psi \partial_x^2 \psi+\frac{1}{2}m^2\psi^2
\end{equation},

 and Lagrangian density

\begin{equation}
\mathcal{L}'=\frac{1}{2}(\partial_t\psi )^2+\frac{1}{2}\psi\partial_x\partial_x \psi -\frac{1}{2}m^2\psi^2
\end{equation}

It is clear that the new Lagrangian give the same equation of motion as the standard equation of motion described previously, because $\mathcal{L' -L}$ is a total divergence. Therefore, the equations of motion generated from Lagrangian alone, are not a sufficient test for choosing the correct Hamiltonian or Lagrangian.

\section{The criteria for choosing the correct Hamiltonian}

The Klein-Gorden equation is the relativistic wave equation for a scalar particle. In the low energy limit, it is reduced to the non-relativistic case. It is then instructive to compare it with the solutions of the non-relativistic Schrodinger wave equation.

Consider the following example of an infinite potential well

\[V(x) = \left\{
\begin{array}{l l}
  0 & \mbox{$0<x<L$}\\
  \infty & \mbox{elsewhere}\\ \end{array} \right. \]

The Schrodinger equation is

\begin{equation}
E_n\phi(x)=-\frac{\partial_x^2\phi(x) }{2m}+V(x)\phi(x)
\end{equation}

Since the particle can not exist inside an infinite potential well, the boundary condition is $\phi=0$ if $x>L$ or $x<0$. The solution is

\begin{eqnarray}
\phi(x)&=&A\sin(k_n x)\\
k_n&=&\frac{n\pi}{L}\\
E_n&=&\frac{k_n^2}{2m}
\end{eqnarray}

Here, $A$ is a constant. $K_n<<m$ so that the relativistic effects can be ignored. The energy density or Hamiltonian density is then
\begin{equation}
\label{sh-de}
H_n=E_n\phi(x)^2=E_n A^2 \sin^2(k_n x)
\end{equation}

The same boundary condition can be applied to the Klein-Gordon equation, which gives the corresponding solution
\begin{eqnarray}
\label{so1}
\psi(x)&=&B \sin(Et+\eta)\sin(k_n x)\\
E&=&\sqrt{m^2+k_n^2}
\end{eqnarray}
where $B$ is a constant. 
We first try the standard Hamiltonian density.

\begin{eqnarray}
\mathcal{H}&=&\frac{B^2}{2} \Big( E^2 \cos^2 (Et+\eta ) \sin^2(k_n x)\nonumber \\
&+&k_n^2\sin^2 (Et+\eta ) \cos^2(k_n x)\nonumber \\
&+&m^2\sin^2 (Et+\eta ) \sin^2(k_n x)\Big)
\end{eqnarray}

This is very different from equation(\ref{sh-de}), and in particular, the energy density at $x=0$ and $x=L$ is not zero.

Now we try the new Hamiltonian density.
\begin{equation}
\mathcal{H}'=\frac{B^2 E^2}{2}\sin^2 (k_n x)
\end{equation}

It is the same as equation (\ref{sh-de}), and we can easily adjust $A$ and $B$ so that $H_n=\mathcal{H}'$.  This implies that $\mathcal{H}'$ is more likely to be the correct Hamiltonian density for a free particle.

\section{Why the new Hamiltonian gives a self-consistent result?}

While it is unclear which Lagrangian or Hamiltonian is {\it a priori} the correct one, we must be able to generate Hamilton's equations of motion directly from the Hamiltonian

\begin{eqnarray}
\label{Hamilton1}
\partial_t \pi&=&-\frac{\partial \mathcal{H}}{\partial{\psi}}\\
\partial_t\psi&=&\frac{\partial \mathcal{H}}{\partial \pi}\\
\pi &=& \frac{\partial \mathcal{L}}{\partial(\partial_t \psi) }=\partial_t \psi
\end{eqnarray}

Following Hamilton's principle, equations (\ref{Hamilton1}) must reproduce the correct equation of motion coming from the Lagrangian. Therefore,

\begin{equation}
\partial_t^2 \psi=-\frac{\partial \mathcal{H}}{\partial{\psi}}=\partial_x^2 \psi -m^2 \psi
\end{equation}

Here, $\partial_x^2\psi-m^2 \psi$ must be $\pi$ independent. Therefore, the Hamiltonian must be in the form

\begin{equation}
 \mathcal{H}_n=\frac{1}{2}(\partial_t \psi)^2 -\int (\partial_x^2 \psi) d\psi+\frac{1}{2}m^2 \psi^2
\end{equation}

Following the calculations in \cite{Morse}, we get

\begin{equation}
\int (\partial_x^2 \psi )d\psi=\frac{1}{2}\psi\partial_x^2 \psi
\end{equation}

Therefore $\mathcal{H}'$ has the same form as $\mathcal{H}_n$ and will give the correct equation of motion. On the other hand, since $\mathcal{H}_n$ is not the same as $\mathcal{H}$,  $\mathcal{H}$ can not generate the correct equation of motion and can not be the correct Hamiltonian density in this case.

\section{The case of general potential}
In general a self-interacting field has an equation of motion

\begin{equation}
\label{motion1}
\partial^2_t \psi-\partial_x^2 \psi+\partial_\psi f(\psi) =0
\end{equation}

Here, $f(\psi)$ is a general function of $\psi$.
Following the same procedure as outlined in the previous sections, one finds that the new Hamiltonian density and Lagrangian density can be written as

\begin{eqnarray}
\mathcal{H}'_g&=&\frac{1}{2}(\partial_t\psi )^2-\frac{1}{2}\psi \partial_x^2 \psi+f(\psi)\\
 \mathcal{L}'_g&=&\frac{1}{2}(\partial_t\psi )^2+\frac{1}{2}\psi \partial_x^2 \psi-f(\psi)\\
\end{eqnarray}

This form is also different from what is generally believed to be the standard Hamiltonian and Lagrangian

\begin{eqnarray}
\mathcal{H}_g&=&\frac{1}{2}(\partial_t\psi )^2+\frac{1}{2}( \partial_x \psi)^2+f(\psi)\\
 \mathcal{L}_g&=&\frac{1}{2}(\partial_t\psi )^2-\frac{1}{2}(\partial_x \psi)^2-f(\psi)
\end{eqnarray}

 Since the difference comes from $\partial_x^2 \psi$ terms, any non-isotropic field will have an energy-momentum distribution different from the original one. The most obvious example is the kink solution. Let's take $Z_2$ symmetry as an example. The $Z_2$ kink is a time independent field. Its energy density depends only on a space coordinate. In this case

\begin{equation}
f(\psi)=\frac{\lambda}{4}(\psi^2-\eta^2)^2
\end{equation}

The equation of motion is

\begin{equation}
\partial^2_t \psi-\partial_x^2 \psi+\lambda (\psi^2-\eta^2)\psi =0
\end{equation}

The $Z_2$-kink solution is

\begin{equation}
\psi=\eta \tanh \Big(\sqrt{\frac{\lambda}{2}} \eta x\Big)
\end{equation}

The new Hamiltonian density or energy density is

\begin{eqnarray}
\mathcal{H}'_g=\frac{\lambda \eta^4}{4} \text{sech}^4 \Big(\sqrt{\frac{\lambda}{2}} \eta x\Big) \cosh\Big(\sqrt{2\lambda} \eta x\Big)
\end{eqnarray}

which is obviously different from the standard one

\begin{equation}
\mathcal{H}_g=\frac{\lambda \eta^4}{2} \text{sech}^4 \Big(\sqrt{\frac{\lambda}{2}} \eta x\Big)
\end{equation}

Since the energy distribution is different in these two cases, the gravitational field of this kink will be different from what is expected. If we were to solve Einstein's equations $G_{\mu \nu} = \kappa T_{\mu \nu}$ in order to find the corresponding metric for this solution, we would need to use the correct energy density $H_g'$, and not the standard one $H_g$.

\section{Conclusions}

In this paper we have argued that it is impossible to guess the Lagrangian and Hamiltonian purely from the equation of motion; extra constraints must be included to fix the choice. As one of the basic requirements, the choice must be consistent with the non-relativistic limit. By imposing this condition we have shown that the standard Hamiltonian density for a scalar field, $\mathcal{H}$, is not the energy density. Instead the correct choice is $\mathcal{H}'$.
The new Hamiltonian $\mathcal{H}'$ also correctly generates Hamilton's equations, which follow from Hamilton's principle. The standard Hamiltonian, $\mathcal{H}$, failed to do this which implies that $\mathcal{H}$ is not always a correct choice. 

We explicitly studied the example of a $Z_2$ kink. As expected, the new energy density is not the same as standard one, which means that its gravitational effect may need to be reconsidered.

The problem for the Hamiltonian and Lagrangian transform in quantum theory is that $(\partial_x \psi)^2$ is treated as a function of $\psi$ and is independent from $\partial_t\psi$. This is not always true, which can be seen from the point of view of a Lorentz boosted observer. After boosting, $\partial_t \psi$ will combine with $\partial_x \psi$. Therefore, there is no reason to believe $\partial_x \psi$ is generally independent from $\partial_t \psi$.

Since scalar fields have been widely studied in particle physics and cosmology the choice of the correct Hamiltonian is very important. The Hamiltonian represents the energy density, so the gravitational effects predicted by these models are not correct. This calls for careful re-examination of the standard results derived in this context.

\begin{acknowledgments}
The author thanks Dejan Stojkovic and Per Sundin for useful discussions.
\end{acknowledgments}

\appendix

\section{general conservation equation}

This is an example of an equation written in the form of the continuity equation:
\begin{eqnarray}
&&\partial_t\Big(\frac{1}{2}(\partial_t\psi )^2+\frac{a}{2}( \partial_x \psi)^2-(1-a)\frac{1}{2}\psi \partial_x^2 \psi+\frac{1}{2}m^2\psi^2 \Big)\nonumber\\
&-&\partial_x\Big((1+a)\frac{1}{2}\partial_x\psi \partial_t \psi-(1-a)\frac{1}{2}\psi\partial_x\partial_t \psi\Big)=0
\end{eqnarray}

Here, $a$ is a constant and

\begin{eqnarray}
&-&\partial_t\Big((1+a)\frac{1}{2}\partial_t\psi \partial_x \psi-(1-a)\frac{1}{2}\psi\partial_x\partial_t \psi \Big)\nonumber \\
&+&\partial_x\Big(\frac{a}{2}(\partial_t\psi )^2-(1-a)\frac{1}{2}\psi \partial_t^2\psi+\frac{1}{2}(\partial_x\psi)^2 -\frac{m^2}{2}\psi^2\Big)=0\nonumber \\
\end{eqnarray}

If we interpret it as the continuity equation, then
\begin{eqnarray}
T^{tt}&=&\frac{1}{2}(\partial_t\psi )^2+a\frac{1}{2}(\partial_x\psi)^2-(1-a)\frac{1}{2}\psi \partial_x^2 \psi+\frac{1}{2}m^2 \psi^2\nonumber\\
\\
T^{tx}&=& T^{xt}=-(1+a)\frac{1}{2}\partial_t\psi \partial_x \psi+(1-a)\frac{1}{2}\psi\partial_x\partial_t \psi\nonumber\\
\\
T^{xx}&=&a\frac{1}{2}(\partial_t\psi )^2-(1-a)\frac{1}{2}\psi \partial_t^2\psi+\frac{1}{2}(\partial_x \psi)^2 -\frac{m^2}{2}\psi^2\nonumber\\
\end{eqnarray}

\end{document}